\documentclass[aps,prl,10pt,twocolumn,superscriptaddress,showpacs]{revtex4-1}
\usepackage[dvipdfmx]{graphicx}
\usepackage{color}

\begin{document}

% --------------- Title --------------------
\title{Exploring a potential energy surface by machine learning for characterizing atomic transport}

% --------------- Authors --------------------
\author{Kenta \surname{Kanamori}}
\thanks{Equally contributed}
\email{kanamori.k.mllab.nit@gmail.com}
\affiliation{Department of Computer Science, Nagoya Institute of Technology, 466-8555, Japan}

\author{Kazuaki \surname{Toyoura}}
\thanks{Equally contributed}
\thanks{Co-corresponding author}
\email{toyoura.kazuaki.5r@kyoto-u.ac.jp}
\affiliation{Department of Materials Science and Engineering, Kyoto University, Kyoto 606-8501, Japan}

\author{Junya \surname{Honda}}
\affiliation{Department of Complexity Science and Engineering, The University of Tokyo, Chiba 277-8561, Japan}
\affiliation{RIKEN Center for Advanced Intelligence Project, Tokyo 103-0027, Japan}

\author{Kazuki \surname{Hattori}}
\affiliation{Department of Materials Science and Engineering, Kyoto University, Kyoto 606-8501, Japan}

\author{Atsuto \surname{Seko}}
\affiliation{Department of Materials Science and Engineering, Kyoto University, Kyoto 606-8501, Japan}
\affiliation{Center for Elements Strategy Initiative for Structure Materials (ESISM), Kyoto University, Kyoto 606-8501, Japan}
\affiliation{JST, PRESTO, Kawaguchi 332-0012, Japan}
\affiliation{Center for Materials Research by Information Integration, National Institute for Materials Science, Tsukuba 305-0047, Japan}

\author{Masayuki \surname{Karasuyama}}
\affiliation{Department of Computer Science, Nagoya Institute of Technology, 466-8555, Japan}
\affiliation{JST, PRESTO, Kawaguchi 332-0012, Japan}
\affiliation{Center for Materials Research by Information Integration, National Institute for Materials Science, Tsukuba 305-0047, Japan}

\author{Kazuki \surname{Shitara}}
\affiliation{Center for Materials Research by Information Integration, National Institute for Materials Science, Tsukuba 305-0047, Japan}
\affiliation{Nanostructures Research Laboratory, Japan Fine Ceramics Center, Nagoya 456-8587, Japan}

\author{Motoki \surname{Shiga}}
\affiliation{Department of Electrical, Electronic and Computer Engineering, Gifu University, Gifu 501-1193, Japan}
\affiliation{JST, PRESTO, Kawaguchi 332-0012, Japan}

\author{Akihide \surname{Kuwabara}}
\affiliation{Nanostructures Research Laboratory, Japan Fine Ceramics Center, Nagoya 456-8587, Japan}
\affiliation{Center for Materials Research by Information Integration, National Institute for Materials Science, Tsukuba 305-0047, Japan}

\author{Ichiro \surname{Takeuchi}}
\thanks{Co-corresponding author}
\email{takeuchi.ichiro@nitech.ac.jp}
\affiliation{Department of Computer Science, Nagoya Institute of Technology, 466-8555, Japan}
\affiliation{RIKEN Center for Advanced Intelligence Project, Tokyo 103-0027, Japan}
\affiliation{Center for Materials Research by Information Integration, National Institute for Materials Science, Tsukuba 305-0047, Japan}

% --------------- Date --------------------
\date{\today}

% --------------- Authors --------------------

\begin{abstract}
We propose a machine-learning method for evaluating the potential barrier governing atomic transport based on the preferential selection of dominant points for the atomic transport.
The proposed method generates numerous random samples of the entire potential energy surface (PES) from a probabilistic Gaussian process model of the PES, which enables defining the likelihood of the dominant points.
%
%The robustness and efficiency of the method are demonstrated on simple model cases of the proton diffusion in $c$-${\rm BaZrO_3}$ and $t$-${\rm LaNbO_4}$, in comparison with a conventional nudge elastic band method.
The robustness and efficiency of the method are demonstrated on a dozen model cases for proton diffusion in oxides, in comparison with a conventional nudge elastic band method.
\end{abstract}

\maketitle

% -------------------- Sec1 --------------------

Atomic transport plays a key role in a variety of phenomena related to physics, chemistry, and materials science.
Concerning the transport of a mobile atom governed by thermally activated processes in a crystal, the kinetics is fully characterized by the entire potential energy surface (PES) of the mobile atom in the host crystal.
The most important region in the entire PES is the {\it optimal path}, which is defined as the lowest-energy path between two global minimum points separated by a lattice translation vector.
Therefore, the optimal path is identical to a valley line, which generally passes through several saddle points.
Based on transition state theory (TST) \cite{laidler1941theory,vineyard1957frequency,toyoura2008first,toyoura2010effects}, the kinetics is determined primarily by the potential barrier of the optimal path, i.e., the difference in potential energy (PE) between the global minimum point and the {\it bottleneck point}, which is defined as the point having the highest PE on the optimal path.

The nudged elastic band (NEB) method \cite{henkelman2000climbing,henkelman2000improved} is a well-established and powerful technique for identifying the optimal path and its energy profile.
The NEB method, however, has a serious drawback in that it requires a given initial trajectory to identify the optimal path. 
When the optimal path consists of several elementary paths, all local minimum points (including the global minimum point) in the entire PES are found in advance, because these points may be the initial and final points of the elementary paths.
For each of the possible initial trajectories derived from the local minimum points, an elementary path is analyzed using the NEB method, and the optimal path formed by some of the elementary paths is finally identified.
Moreover, physical and chemical prior knowledge, e.g., ionic radii, chemical bonding states, and electrostatic interaction, has usually been used to obtain local minimum points and initial trajectories.
However, the excessive dependence on the prior knowledge may cause us to miss a key elementary path.
Thus, a robust and efficient alternative method to the NEB-based analysis is desired. 

In the present study, we propose a machine-learning (ML) method for robustly and efficiently estimating the potential barrier of the optimal path. 
The basic strategy is to focus only on finding the two dominant points, i.e., the global minimum and bottleneck points. 
To this end, a probabilistic Gaussian process (GP) model \cite{williams2006gaussian,stein2012interpolation} of the PES is introduced, and is iteratively updated using the first-principles potential energies (PEs) computed at the selected points based on the {\it uncertainties} of the two dominant points within a Bayesian optimization (BO)-like framework. 
The uncertainties are obtained from the probabilistic PES model and optimal path searches by a dynamic programming (DP)-base algorithm \cite{bellman2013dynamic,korte2012combinatorial,kaibel2006bottleneck}.

Although probabilistic approaches based on BO were previously used in several materials informatics studies \cite{PhysRevLett.115.205901,toyoura2016machine,xue2016accelerated,Kiyoharae1600746,PhysRevB.95.144110,2017arXiv170809274T}, these existing methods cannot be directly used in the present case.
A kernel-based support vector machine (SVM) technique was also reported for finding transition pathways  \cite{pozun2012optimizing}, in which all saddle points on all elementary paths are explored, not only the single bottleneck point on the optimal path, as in the current problem.
The intrinsic difficulty of the problem considered herein is that the optimal path and its bottleneck point are, by definition, found after acquiring complete information about the entire PES.  
The basic strategy for overcoming this difficulty is to randomly generate multiple PES samples according to a probabilistic GP model and to identify the optimal path for each sample using a DP-based algorithm. 
This enables collections of the global minimum and bottleneck points to be obtained, and these collections are considered to be distributions representing the uncertainties of these two dominant points. 
In order to implement this concept, the GP model, the DP algorithm, and the BO framework are properly extended and effectively combined.
Note that the proposed ML method is, in principle, applicable to any kind of atomic transport phenomena governed by multiple mobile atoms or ions.
Furthermore, other phenomena governed by thermally activated processes, e.g., phase transitions and chemical reactions, are also considered as applications, where both initial and final states can be given in a configuration space.

%The robustness and efficiency of the proposed ML method are demonstrated by comparing the proposed method with the NEB-based analysis \cite{henkelman2000climbing,henkelman2000improved} for simple model cases. 
The robustness and efficiency of the proposed ML method are demonstrated for a dozen model cases for proton diffusion in host oxides with a variety of crystal structures. 
%
%These cases are the proton diffusion in two well-known proton-conducting oxides, i.e., barium zirconate with the cubic perovskite structure ($c$-${\rm BaZrO_3}$) \cite{iwahara1993protonic,munch2000proton,bjorketun2007effect,sundell2007density} and lanthanum niobate with a tetragonal scheelite structure ($t$-${\rm LaNbO_4}$) \cite{haugsrud2006proton1,haugsrud2006proton2}.
In the present paper, the results for two well-known proton-conducting oxides, i.e., barium zirconate with a cubic perovskite structure ($c$-${\rm BaZrO_3}$) \cite{iwahara1993protonic,munch2000proton,bjorketun2007effect,sundell2007density} and lanthanum niobate with a tetragonal scheelite structure ($t$-${\rm LaNbO_4}$) \cite{haugsrud2006proton1,haugsrud2006proton2}, are taken as the primary typical examples, to show the applicability of the proposed ML method to both isotropic and anisotropic diffusivities.
Due to space limitations, the results in the other cases are summarized in Supplemental Materials F, which are also of importance for demonstrating the general applicability of the proposed method. 
%We evaluated the entire PES of a proton in $c$-${\rm BaZrO_3}$ in a recent study \cite{toyoura2016machine}, which is used to validate the proposed ML method and to check the performance.
%
%On the other hand, the entire PES of $t$-${\rm LaNbO_4}$, which has a lower crystallographic symmetry, has not yet been evaluated and is used to verify the applicability to anisotropic diffusivity characterized by two types of optimal paths in the crystal.

% -------------------- Sec2 --------------------

The outline of the proposed ML method is explained hereinafter.
First, a fine grid for computing PEs is introduced in the conventional unit cell of a host crystal. 
By fully exploiting the crystallographic symmetry, we consider only the irreducible grid points, which correspond to the grid points in the asymmetric unit. 
As the initialization process, the PEs at several grid points are computed, which are randomly selected from the asymmetric unit (10 grid points in the present study).
The current problem is to efficiently obtain the PEs at the global minimum and bottleneck points, $E^{\rm min}$ and $E^{\rm btl}$, respectively, by as few PE computations as possible.
The PE difference between the global minimum and bottleneck points is denoted here by $\Delta E \: (= E^{\rm btl} - E^{\rm min})$.
The mathematical formulation of the problem setting is detailed in Supplemental Materials A.

\begin{figure}[tbp]
\begin{center}
\includegraphics[width=\linewidth]{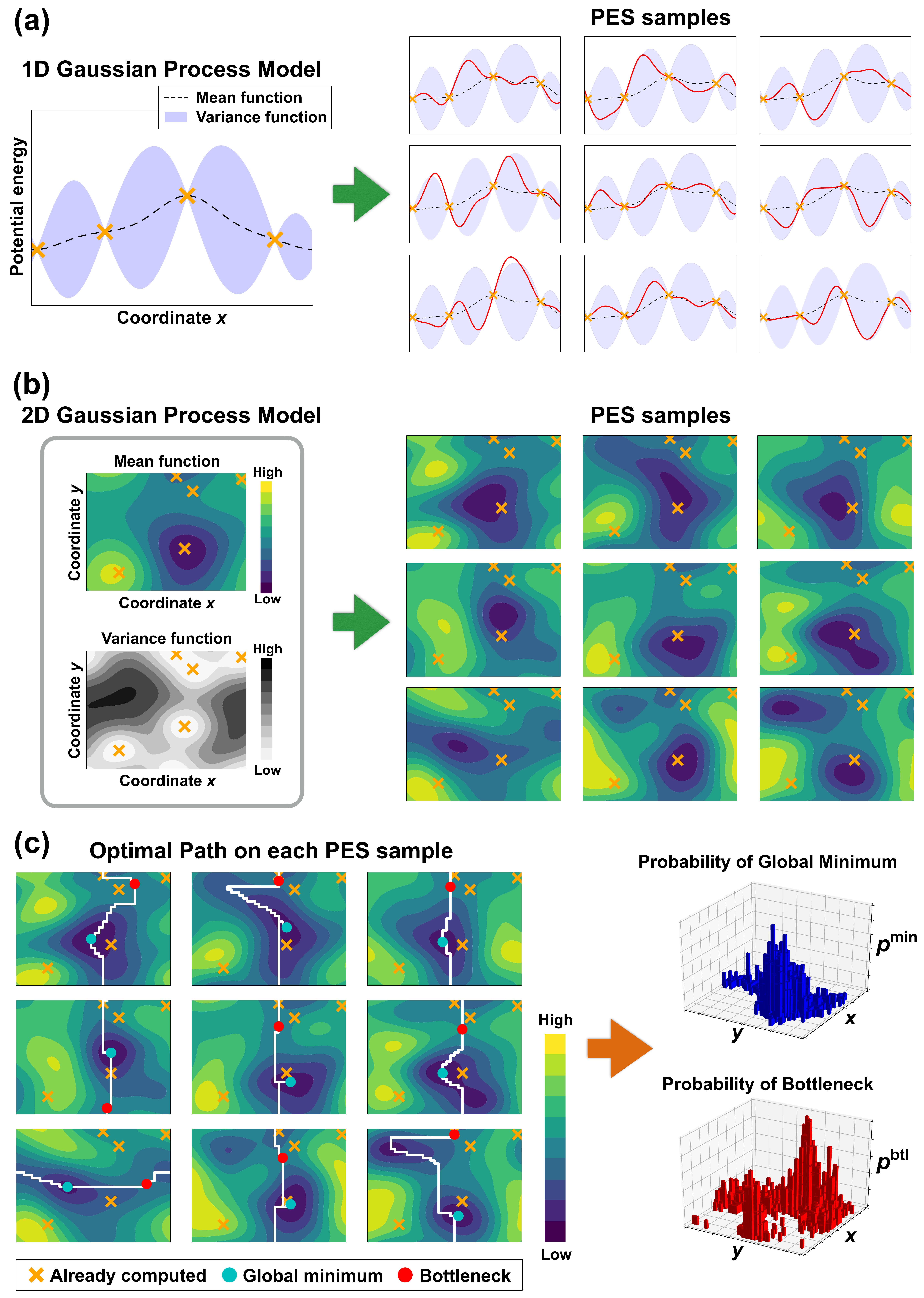}
\caption{
Schematic diagram of random PES samples for (a) synthetic 1D PES examples, (b) synthetic 2D PES examples, and (c) the optimal paths in the 2D PES examples.
In (a), the black curve and blue area, respectively, represent the mean and the variance functions of the GP-based probabilistic PES model.
On the right, nine randomly generated PES samples are indicated by red curves.
In (b), the color and the gray-scale contour plots, respectively, on the left represent the mean and the variance functions of the GP-based probabilistic PES model.
On the right, nine randomly generated PES samples are indicated by color contour plots.
In the left-hand plot of (c), the white curve in each plot indicates the optimal path identified by the DP-based path search algorithm.
On the right, the blue and red histograms, respectively, represent the collections of the global minimum and bottleneck points obtained from 1,000 random PES samples and the optimal path search for each sample.
These collections of points can be interpreted as the probability distributions of the two dominant points and are used for selecting the next PE computation point.
}
\label{fig:method}
\end{center}
\end{figure}

The proposed ML method solves the problem by iterating the following three steps (steps I, II, and III).
In step I, a probabilistic PES model is constructed by GP using the already-computed PEs obtained by first-principles calculations in the earlier iterations, and randomly generating a large number of PES samples from the probabilistic model (See Supplemental Materials B for details).
Figure \ref{fig:method}(a) shows an example of step I for a synthetic one-dimensional (1D) case, where nine PES samples are shown as typical examples.
These PES samples fluctuate around the mean function of the probabilistic model reflecting its uncertainty.
An example of the synthetic 2D case is also shown in Fig. \ref{fig:method}(b).
In the present study, a thousand randomized PES samples are actually generated in the 3D configuration space. 

In step II, the optimal path is identified in each of the randomized PES samples using a DP-based algorithm.
Figure \ref{fig:method} (c) shows the optimal path in each of the nine randomized PES samples.
The optimal path depends on the randomness of the PES samples, resulting in the uncertainties of the global minimum and bottleneck points as shown in Fig. \ref{fig:method} (c).
The uncertainties are represented by the probabilities of the $i$-th point to be the global minimum and bottleneck points in the multiple PES samples, $p^{\rm min}_i$ and $p^{\rm btl}_i$, respectively.
If atomic transport is characterized by multiple optimal paths due to the low symmetry of a host crystal (e.g., $t$-${\rm LaNbO_4}$), subsequent optimal paths with a higher $\Delta E$ can also be identified one by one in the same manner.
The details of the path search algorithm based on DP and the extension for atomic transport with biaxial or triaxial anisotropy are presented in Supplemental Materials C.

In step III, the next grid point, which is likely to be either the global minimum point or a bottleneck point, is selected using a BO-like approach, and the PE at the selected point is computed using first-principles calculations.
The next point is determined as the point that maximizes $(p^{\rm min}_i + p^{\rm btl}_i)\sigma^2_i$, where $\sigma^2_i$ represents the uncertainty of the PE at the $i$-th point. (Details are provided in Supplemental Materials D.) 

\begin{figure}[tbp]
\begin{center}
\includegraphics[width=\linewidth]{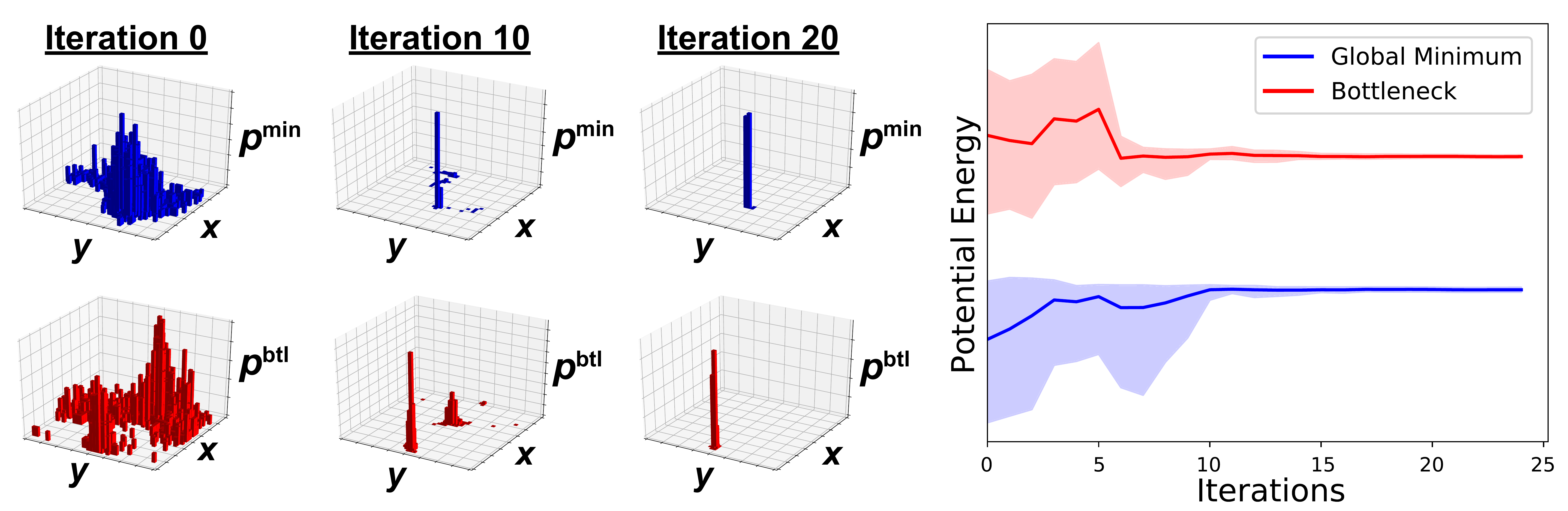}
\caption{
Schematic diagrams representing how the uncertainties of the two dominant points decrease in the 2D examples in Fig. \ref{fig:method}(c).
The histograms in the same form as Fig. \ref{fig:method}(c) at the initial, $10$-th, and $20$-th iterations are plotted on the left.
The mean (curves) and the variances (shaded areas) representing how the uncertainties of the global minimum point (blue) and the bottleneck point (red) decrease through iterations are plotted on the right.
These plots suggest that the estimations of the locations of the two dominant points converge to the true locations as the iterations proceed.
}
\label{fig:method2}
\end{center}
\end{figure}

Figure \ref{fig:method2} shows the probability distributions of the global minimum and bottleneck points, $p^{\rm min}_i$ and $p^{\rm btl}_i$, in the initial, 10-th and 20-th iterations. 
These distributions have large variance in earlier iterations, but the peaks of the distributions gradually converge to the true global minimum and bottleneck points. 
The variances of the PEs at the global minimum and bottleneck points also converge with iterations (as shown on the right).
Hence they are used as a stopping criterion for an optimal path search ($\text{the variance } < 10^{-6}$ eV in the present study).

% -------------------- Sec3 --------------------

%First-principles calculations in both model systems were based on the projector augmented wave (PAW) method as implemented in the VASP code \cite{blochl1994projector,kresse1993ab,kresse1996efficiency,kresse1999ultrasoft} (See Refs. \cite{toyoura2016machine} and \cite{fjeld2010proton} for the detailed computational conditions).
The PEs at each grid point in both model systems were computed from first principles based on the projector augmented wave (PAW) method as implemented in the VASP code \cite{blochl1994projector,kresse1993ab,kresse1996efficiency,kresse1999ultrasoft}. (See Refs. \cite{toyoura2016machine} and \cite{fjeld2010proton} for the detailed computational conditions.)
The lattice relaxation around the proton was taken into account by structural optimization with only the proton position fixed at the grid point.
We introduced a 40$\times$40$\times$40 grid for $c$-${\rm BaZrO_3}$ and a 20$\times$20$\times$40 grid for $t$-${\rm LaNbO_4}$, which generated 1,768 and 1,010 irreducible grid points, respectively.

\begin{figure}[tbp]
\begin{center}
\includegraphics[width=\linewidth]{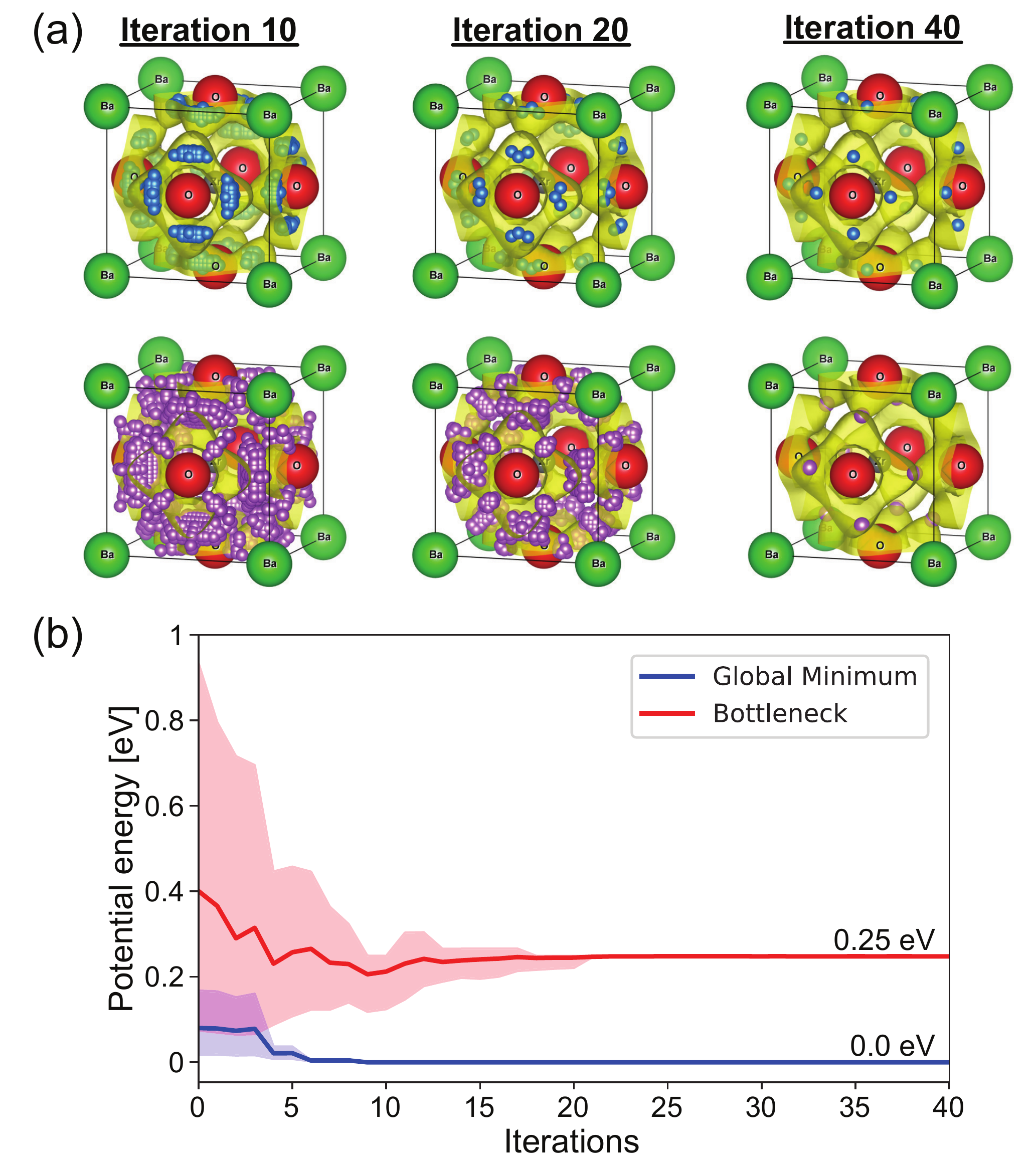}
\caption{
(a) Global minimum (cyan) and bottleneck (purple) points in the 1,000 randomized proton PES samples in the 10th, 20th, and 40th iterations in the $c$-$\rm {BaZrO}_3$ crystal.
(b) Mean and variance profiles of the PEs at the global minimum and bottleneck points in each iteration. The PE values are shown with reference to the true PE at the global minimum point.
}
\label{fig:result1}
\end{center}
\end{figure}

The first model case is the proton diffusion in $c$-${\rm BaZrO_3}$, in which the entire PES is known and the isotropic proton diffusivity is characterized by a single optimal path.
The yellow surface shown in Fig. \ref{fig:result1} (a) denotes the PE isosurface in the $c$-${\rm BaZrO_3}$ crystal (PE level: 0.3 eV with reference to the global minimum point).
The low PE region surrounded by the isosurface is located around oxygen ions, forming the proton rotational orbits.
These rotational orbits overlap adjacent proton rotational orbits, corresponding to the hopping paths between oxygen ions.
The two elementary paths, i.e., the rotation and hopping paths, form a 3D proton-conducting network throughout the crystal lattice.
Their calculated potential barriers are 0.18 and 0.25 eV, respectively.
Therefore, the bottleneck point corresponds to the saddle point for the proton hopping, not for the proton rotation.

The purple and cyan dots in Fig. \ref{fig:result1} (a) indicate grid points predicted at least once as the global minimum and bottleneck points, respectively, in the 10th, 20th, and 40th iterations.
Figure \ref{fig:result1} (b) shows the mean and variance of the predicted PEs for the global minimum and bottleneck points as a function of the number of the iterations.
At the beginning (iteration 10), the predicted global minimum and bottleneck points are scattering, and the predicted PEs at the global minimum and bottleneck points deviate from the true values (0 eV and 0.25 eV, respectively) with large variances.
As the iterations proceed, the number of grid points predicted as the global minimum and bottleneck points gradually decreases, and the predicted PEs approach the true values while their variances become smaller, until finally converging to the true points and PEs with tiny variances (iteration 40).
The proposed ML method found the true potential barrier after 40 iterations, i.e., the proton diffusivity in this crystal is completely characterized by first-principles calculations for 50 points, including 10 points that are randomly selected in the initialization process.

\begin{figure}[tbp]
\begin{center}
\includegraphics[width=\linewidth]{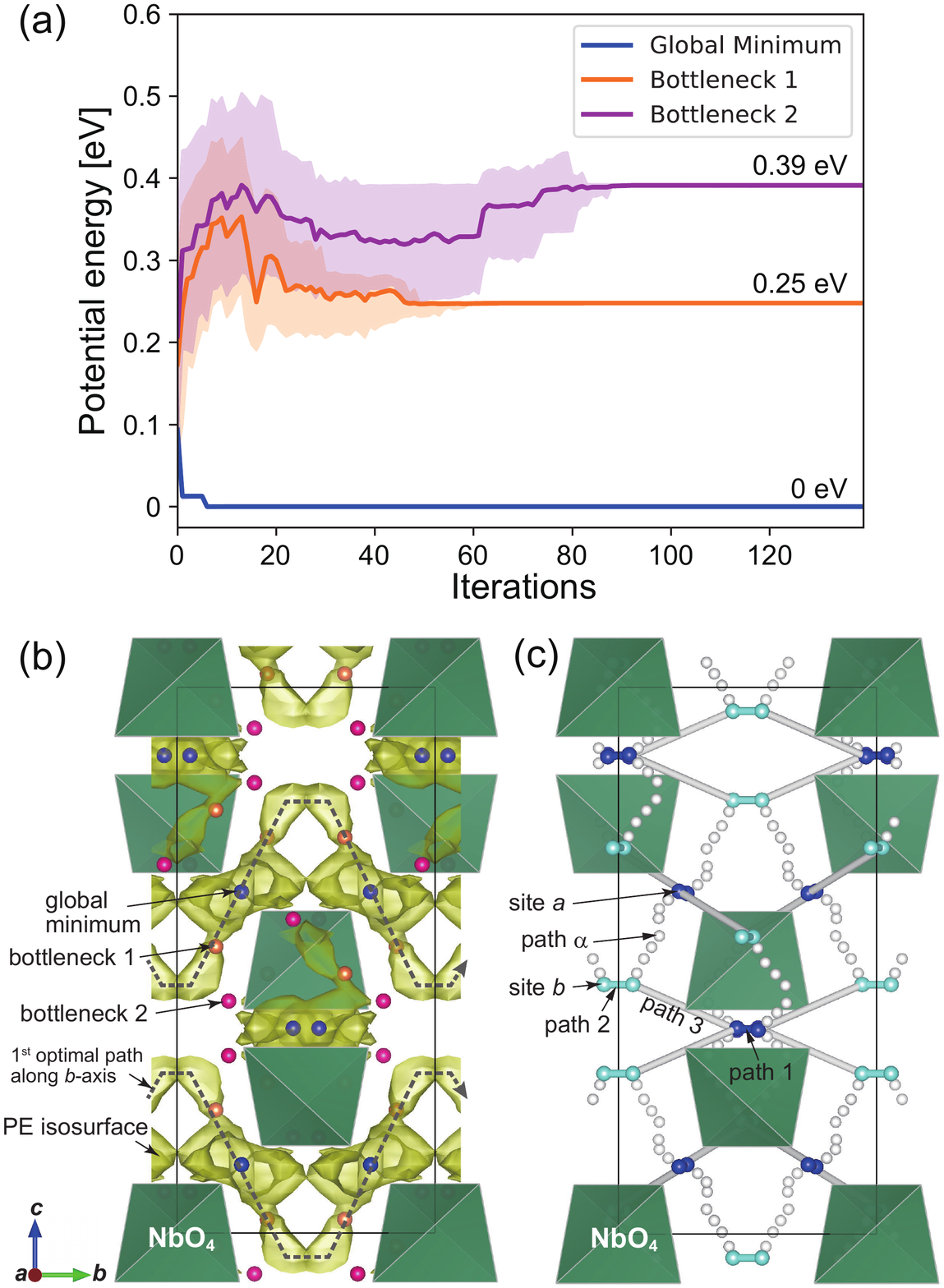}
\caption{(a) Mean and variance profiles of the PEs at the global minimum and two bottleneck points in $t$-$\rm {LaNbO}_4$. The PE values are shown with reference to the true PE at the global minimum point. (b) Global minimum and two bottleneck points along with the proton PE isosurface (PE level: 0.25 eV) in the PES predicted by the probabilistic model. (c) Reported global and local minimum points (sites $a$ and $b$) and paths 1$\sim$3 defined in Ref. \cite{fjeld2010proton}. The white spheres indicate the trajectory of path $\alpha$ additionally calculated by the NEB method in the present study.}
\label{fig:result2}
\end{center}
\end{figure}

The proposed ML method can readily be extended to anisotropic diffusivity in a crystal with a low symmetry.
The anisotropic proton diffusivity in $t$-${\rm LaNbO_4}$ is here taken as another model case, in which the proton diffusivity is different between the $ab$-plane and the $c$-axis from the crystallographic point of view.
In this case, the global minimum and bottleneck points on the optimal path are first explored.
Subsequently, the bottleneck point on the second optimal path, which is defined as the second-lowest-energy path that is linearly independent of the first optimal path, is additionally explored using an extension of the proposed method (See Supplemental Materials C.2).
Figure \ref{fig:result2} (a) shows the mean and variance profiles of the PEs at the global minimum and two bottleneck points on the first and second optimal paths.
They correspond to the optimal paths in the $ab$-plane and along the $c$-axis, respectively, showing the potential barriers of $\Delta E_{\rm ab} = 0.25$ eV and $\Delta E_{\rm c} = 0.39$ eV.

The predicted potential barriers are, however, inconsistent with the reported values ($\Delta E_{\rm ab} = 0.41$ eV and $\Delta E_{\rm c} = 0.74$ eV) in the previous study using the NEB-based analysis \cite{fjeld2010proton}. 
The origin of the difference is missing a key elementary path through the bottleneck point (bottleneck 1) in the previous study.
Figure \ref{fig:result2} (b) shows the global minimum and two bottleneck points along with the proton PE isosurface (PE level: 0.25 eV) in the PES predicted by the probabilistic model in the final iteration.
The first optimal path runs in the region surrounded by the PE isosurface, which forms equivalent 1D long-range paths along the $a$- and $b$-axes \footnote{The 1D long-range paths along the $a$- and $b$-axes are alternatively arranged in the $c$-axis direction due to the four-fold screw symmetry. Note that the broken lines in Fig. \ref{fig:result2}(b) show only the paths along the $b$-axis.}.
On the other hand, Fig. \ref{fig:result2} (c) shows the global and local minimum points (sites $a$ and $b$) and 1D long-range paths consisting of paths 1, 2 and 3 found by the NEB-based analysis in Ref \cite{fjeld2010proton}.
The white spheres denote the key elementary path (path $\alpha$) additionally calculated by the NEB method in the present study.
The converged trajectory is in line with the predicted first optimal path, the calculated potential barrier (0.25 eV) of which is equal to the predicted PE at bottleneck 1.
Hence paths 1, 2, and $\alpha$ should form the 1D long-range path along the $a$- and $b$-axes, in which path $\alpha$ plays a key role in determining the proton diffusivity.

Concerning the second optimal path along the $c$-axis, the bottleneck point (bottleneck 2) is located on reported path 3, which connects adjacent first optimal paths and enables protons to migrate along the $c$-axis.
Since path 3 is rate-determining in the second optimal path, the predicted $\Delta E_{\rm c}$ in the present study is comparable to the reported potential barrier of path 3. 
The slight difference is probably an artifact due to the interval of the grid points introduced in the proposed method.
Thus, the proton diffusivity in $t$-${\rm LaNbO_4}$ is correctly updated by the proposed ML method in the present study.

Note that the proposed ML method shows high computational efficiency for identifying both global minimum and bottleneck points.
This feature is emphasized in applications to atomic diffusion in low-symmetry crystals, which requires covering all initial trajectories connecting a number of local minimum points for finding the optimal path in the NEB-based analysis.
%
%Actually, in Ref. \cite{fjeld2010proton}, a key proton migration path could not be found after roughly evaluating the entire PES, performing five structural optimizations for determining the local minimum points, and performing 21 careful NEB calculations.
%
%On the other hand, the proposed method requires only a hundred PE computations including the initialization process to identify the two optimal paths and their  $\Delta E$.
%
Actually, in Ref. \cite{fjeld2010proton}, a key proton migration path could not be found after roughly evaluating the entire PES (1,000 single point calculations), performing five structural optimizations for determining the local minimum points, and performing 21 careful NEB calculations.
The total computational cost is approximately equal to 200 PE computations, which is twice that of the ML method proposed in the present study.
The proposed method also works efficiently for the proton diffusions in other host oxides (See Supplementary Materials F), which implies the general applicability to any atomic transport.
Note that the efficiency increases as the symmetry of the host oxide crystal decreases, which indicates that the proposed method is more efficient and useful than the NEB method for identifying complicated atomic transport in low-symmetry crystals.

% -------------------- Sec4 --------------------

We have proposed an ML-based method based on the extended frameworks of GP, DP, and BO in order to automatically analyze the atomic transport of interest.
The proposed method significantly accelerates the identification of dominant points of the optimal migration path by preferentially performing the PE computations at points with a high likelihood of being dominant points.
The results of the demonstration studies also indicate that the proposed method is expected to be particularly useful for investigating the atomic diffusion in complicated low-symmetry crystals without missing key migration elementary paths.
The proposed ML method should therefore be an alternative for the conventional NEB-based analysis, as a more robust, efficient, and realistic method.

% -------------------- acknowledgement --------------------

\subsection*{Acknowledgement}
We appreciate for the insightful discussion by S. Nakajima. This work was partially supported by grants from the Japanese Ministry of Education, Culture, Sports, Science and Technology to K.T. (17H04948, 25106002), J.H. (16H00881), A.S. (15H04116), M.K. (17H04694, 16H06538), M.S. (16H00736, 16H02866), and I.T. (17H00758, 16H06538), and JST PRESTO to A.S. (Grant Number JPMJPR15N7), M.K. (Grant Number JPMJPR15N2), M.S. (Grant Number JPMJPR16N6), JST CREST to I.T. (Grant Number JPMJCR1302, JPMJCR1502), RIKEN Center for Advanced Intelligence Project to J.H and I.T., and by JST support program for starting up innovation-hub on materials research by information integration initiative to A.S., M.K., K.S., A.K., and I.T.


\begin{thebibliography}{30}%
\makeatletter
\providecommand \@ifxundefined [1]{%
 \@ifx{#1\undefined}
}%
\providecommand \@ifnum [1]{%
 \ifnum #1\expandafter \@firstoftwo
 \else \expandafter \@secondoftwo
 \fi
}%
\providecommand \@ifx [1]{%
 \ifx #1\expandafter \@firstoftwo
 \else \expandafter \@secondoftwo
 \fi
}%
\providecommand \natexlab [1]{#1}%
\providecommand \enquote  [1]{``#1''}%
\providecommand \bibnamefont  [1]{#1}%
\providecommand \bibfnamefont [1]{#1}%
\providecommand \citenamefont [1]{#1}%
\providecommand \href@noop [0]{\@secondoftwo}%
\providecommand \href [0]{\begingroup \@sanitize@url \@href}%
\providecommand \@href[1]{\@@startlink{#1}\@@href}%
\providecommand \@@href[1]{\endgroup#1\@@endlink}%
\providecommand \@sanitize@url [0]{\catcode `\\12\catcode `\$12\catcode
  `\&12\catcode `\#12\catcode `\^12\catcode `\_12\catcode `\%12\relax}%
\providecommand \@@startlink[1]{}%
\providecommand \@@endlink[0]{}%
\providecommand \url  [0]{\begingroup\@sanitize@url \@url }%
\providecommand \@url [1]{\endgroup\@href {#1}{\urlprefix }}%
\providecommand \urlprefix  [0]{URL }%
\providecommand \Eprint [0]{\href }%
\providecommand \doibase [0]{http://dx.doi.org/}%
\providecommand \selectlanguage [0]{\@gobble}%
\providecommand \bibinfo  [0]{\@secondoftwo}%
\providecommand \bibfield  [0]{\@secondoftwo}%
\providecommand \translation [1]{[#1]}%
\providecommand \BibitemOpen [0]{}%
\providecommand \bibitemStop [0]{}%
\providecommand \bibitemNoStop [0]{.\EOS\space}%
\providecommand \EOS [0]{\spacefactor3000\relax}%
\providecommand \BibitemShut  [1]{\csname bibitem#1\endcsname}%
\let\auto@bib@innerbib\@empty
%</preamble>
\bibitem [{\citenamefont {Laidler}\ and\ \citenamefont
  {Eyring}(1941)}]{laidler1941theory}%
  \BibitemOpen
  \bibfield  {author} {\bibinfo {author} {\bibfnamefont {K.}~\bibnamefont
  {Laidler}}\ and\ \bibinfo {author} {\bibfnamefont {H.}~\bibnamefont
  {Eyring}},\ }\href@noop {} {\bibfield  {journal} {\bibinfo  {journal}
  {McGraw-Hill Book Company}\ } (\bibinfo {year} {1941})}\BibitemShut {NoStop}%
\bibitem [{\citenamefont {Vineyard}(1957)}]{vineyard1957frequency}%
  \BibitemOpen
  \bibfield  {author} {\bibinfo {author} {\bibfnamefont {G.~H.}\ \bibnamefont
  {Vineyard}},\ }\href@noop {} {\bibfield  {journal} {\bibinfo  {journal}
  {Journal of Physics and Chemistry of Solids}\ }\textbf {\bibinfo {volume}
  {3}},\ \bibinfo {pages} {121} (\bibinfo {year} {1957})}\BibitemShut {NoStop}%
\bibitem [{\citenamefont {Toyoura}\ \emph {et~al.}(2008)\citenamefont
  {Toyoura}, \citenamefont {Koyama}, \citenamefont {Kuwabara}, \citenamefont
  {Oba},\ and\ \citenamefont {Tanaka}}]{toyoura2008first}%
  \BibitemOpen
  \bibfield  {author} {\bibinfo {author} {\bibfnamefont {K.}~\bibnamefont
  {Toyoura}}, \bibinfo {author} {\bibfnamefont {Y.}~\bibnamefont {Koyama}},
  \bibinfo {author} {\bibfnamefont {A.}~\bibnamefont {Kuwabara}}, \bibinfo
  {author} {\bibfnamefont {F.}~\bibnamefont {Oba}}, \ and\ \bibinfo {author}
  {\bibfnamefont {I.}~\bibnamefont {Tanaka}},\ }\href@noop {} {\bibfield
  {journal} {\bibinfo  {journal} {Physical Review B}\ }\textbf {\bibinfo
  {volume} {78}},\ \bibinfo {pages} {214303} (\bibinfo {year}
  {2008})}\BibitemShut {NoStop}%
\bibitem [{\citenamefont {Toyoura}\ \emph {et~al.}(2010)\citenamefont
  {Toyoura}, \citenamefont {Koyama}, \citenamefont {Kuwabara},\ and\
  \citenamefont {Tanaka}}]{toyoura2010effects}%
  \BibitemOpen
  \bibfield  {author} {\bibinfo {author} {\bibfnamefont {K.}~\bibnamefont
  {Toyoura}}, \bibinfo {author} {\bibfnamefont {Y.}~\bibnamefont {Koyama}},
  \bibinfo {author} {\bibfnamefont {A.}~\bibnamefont {Kuwabara}}, \ and\
  \bibinfo {author} {\bibfnamefont {I.}~\bibnamefont {Tanaka}},\ }\href@noop {}
  {\bibfield  {journal} {\bibinfo  {journal} {The Journal of Physical Chemistry
  C}\ }\textbf {\bibinfo {volume} {114}},\ \bibinfo {pages} {2375} (\bibinfo
  {year} {2010})}\BibitemShut {NoStop}%
\bibitem [{\citenamefont {Henkelman}\ \emph {et~al.}(2000)\citenamefont
  {Henkelman}, \citenamefont {Uberuaga},\ and\ \citenamefont
  {J{\'o}nsson}}]{henkelman2000climbing}%
  \BibitemOpen
  \bibfield  {author} {\bibinfo {author} {\bibfnamefont {G.}~\bibnamefont
  {Henkelman}}, \bibinfo {author} {\bibfnamefont {B.~P.}\ \bibnamefont
  {Uberuaga}}, \ and\ \bibinfo {author} {\bibfnamefont {H.}~\bibnamefont
  {J{\'o}nsson}},\ }\href@noop {} {\bibfield  {journal} {\bibinfo  {journal}
  {The Journal of chemical physics}\ }\textbf {\bibinfo {volume} {113}},\
  \bibinfo {pages} {9901} (\bibinfo {year} {2000})}\BibitemShut {NoStop}%
\bibitem [{\citenamefont {Henkelman}\ and\ \citenamefont
  {J{\'o}nsson}(2000)}]{henkelman2000improved}%
  \BibitemOpen
  \bibfield  {author} {\bibinfo {author} {\bibfnamefont {G.}~\bibnamefont
  {Henkelman}}\ and\ \bibinfo {author} {\bibfnamefont {H.}~\bibnamefont
  {J{\'o}nsson}},\ }\href@noop {} {\bibfield  {journal} {\bibinfo  {journal}
  {The Journal of chemical physics}\ }\textbf {\bibinfo {volume} {113}},\
  \bibinfo {pages} {9978} (\bibinfo {year} {2000})}\BibitemShut {NoStop}%
\bibitem [{\citenamefont {Williams}\ and\ \citenamefont
  {Rasmussen}(2006)}]{williams2006gaussian}%
  \BibitemOpen
  \bibfield  {author} {\bibinfo {author} {\bibfnamefont {C.~K.}\ \bibnamefont
  {Williams}}\ and\ \bibinfo {author} {\bibfnamefont {C.~E.}\ \bibnamefont
  {Rasmussen}},\ }\href@noop {} {\bibfield  {journal} {\bibinfo  {journal} {the
  MIT Press}\ }\textbf {\bibinfo {volume} {2}},\ \bibinfo {pages} {4} (\bibinfo
  {year} {2006})}\BibitemShut {NoStop}%
\bibitem [{\citenamefont {Stein}(2012)}]{stein2012interpolation}%
  \BibitemOpen
  \bibfield  {author} {\bibinfo {author} {\bibfnamefont {M.~L.}\ \bibnamefont
  {Stein}},\ }\href@noop {} {\emph {\bibinfo {title} {Interpolation of spatial
  data: some theory for kriging}}}\ (\bibinfo  {publisher} {Springer Science \&
  Business Media},\ \bibinfo {year} {2012})\BibitemShut {NoStop}%
\bibitem [{\citenamefont {Bellman}(2013)}]{bellman2013dynamic}%
  \BibitemOpen
  \bibfield  {author} {\bibinfo {author} {\bibfnamefont {R.}~\bibnamefont
  {Bellman}},\ }\href@noop {} {\emph {\bibinfo {title} {Dynamic programming}}}\
  (\bibinfo  {publisher} {Courier Corporation},\ \bibinfo {year}
  {2013})\BibitemShut {NoStop}%
\bibitem [{\citenamefont {Korte}\ \emph {et~al.}(2012)\citenamefont {Korte},
  \citenamefont {Vygen}, \citenamefont {Korte},\ and\ \citenamefont
  {Vygen}}]{korte2012combinatorial}%
  \BibitemOpen
  \bibfield  {author} {\bibinfo {author} {\bibfnamefont {B.}~\bibnamefont
  {Korte}}, \bibinfo {author} {\bibfnamefont {J.}~\bibnamefont {Vygen}},
  \bibinfo {author} {\bibfnamefont {B.}~\bibnamefont {Korte}}, \ and\ \bibinfo
  {author} {\bibfnamefont {J.}~\bibnamefont {Vygen}},\ }\href@noop {} {\emph
  {\bibinfo {title} {Combinatorial optimization}}},\ Vol.~\bibinfo {volume}
  {2}\ (\bibinfo  {publisher} {Springer},\ \bibinfo {year} {2012})\BibitemShut
  {NoStop}%
\bibitem [{\citenamefont {Kaibel}\ and\ \citenamefont
  {Peinhardt}(2006)}]{kaibel2006bottleneck}%
  \BibitemOpen
  \bibfield  {author} {\bibinfo {author} {\bibfnamefont {V.}~\bibnamefont
  {Kaibel}}\ and\ \bibinfo {author} {\bibfnamefont {M.~A.}\ \bibnamefont
  {Peinhardt}},\ }\href@noop {} {\emph {\bibinfo {title} {On the bottleneck
  shortest path problem}}}\ (\bibinfo  {publisher} {Konrad-Zuse-Zentrum f{\"u}r
  Informationstechnik},\ \bibinfo {year} {2006})\BibitemShut {NoStop}%
\bibitem [{\citenamefont {Seko}\ \emph {et~al.}(2015)\citenamefont {Seko},
  \citenamefont {Togo}, \citenamefont {Hayashi}, \citenamefont {Tsuda},
  \citenamefont {Chaput},\ and\ \citenamefont
  {Tanaka}}]{PhysRevLett.115.205901}%
  \BibitemOpen
  \bibfield  {author} {\bibinfo {author} {\bibfnamefont {A.}~\bibnamefont
  {Seko}}, \bibinfo {author} {\bibfnamefont {A.}~\bibnamefont {Togo}}, \bibinfo
  {author} {\bibfnamefont {H.}~\bibnamefont {Hayashi}}, \bibinfo {author}
  {\bibfnamefont {K.}~\bibnamefont {Tsuda}}, \bibinfo {author} {\bibfnamefont
  {L.}~\bibnamefont {Chaput}}, \ and\ \bibinfo {author} {\bibfnamefont
  {I.}~\bibnamefont {Tanaka}},\ }\href {\doibase
  10.1103/PhysRevLett.115.205901} {\bibfield  {journal} {\bibinfo  {journal}
  {Phys. Rev. Lett.}\ }\textbf {\bibinfo {volume} {115}},\ \bibinfo {pages}
  {205901} (\bibinfo {year} {2015})}\BibitemShut {NoStop}%
\bibitem [{\citenamefont {Toyoura}\ \emph {et~al.}(2016)\citenamefont
  {Toyoura}, \citenamefont {Hirano}, \citenamefont {Seko}, \citenamefont
  {Shiga}, \citenamefont {Kuwabara}, \citenamefont {Karasuyama}, \citenamefont
  {Shitara},\ and\ \citenamefont {Takeuchi}}]{toyoura2016machine}%
  \BibitemOpen
  \bibfield  {author} {\bibinfo {author} {\bibfnamefont {K.}~\bibnamefont
  {Toyoura}}, \bibinfo {author} {\bibfnamefont {D.}~\bibnamefont {Hirano}},
  \bibinfo {author} {\bibfnamefont {A.}~\bibnamefont {Seko}}, \bibinfo {author}
  {\bibfnamefont {M.}~\bibnamefont {Shiga}}, \bibinfo {author} {\bibfnamefont
  {A.}~\bibnamefont {Kuwabara}}, \bibinfo {author} {\bibfnamefont
  {M.}~\bibnamefont {Karasuyama}}, \bibinfo {author} {\bibfnamefont
  {K.}~\bibnamefont {Shitara}}, \ and\ \bibinfo {author} {\bibfnamefont
  {I.}~\bibnamefont {Takeuchi}},\ }\href@noop {} {\bibfield  {journal}
  {\bibinfo  {journal} {Physical Review B}\ }\textbf {\bibinfo {volume} {93}},\
  \bibinfo {pages} {054112} (\bibinfo {year} {2016})}\BibitemShut {NoStop}%
\bibitem [{\citenamefont {Xue}\ \emph {et~al.}(2016)\citenamefont {Xue},
  \citenamefont {Balachandran}, \citenamefont {Hogden}, \citenamefont
  {Theiler}, \citenamefont {Xue},\ and\ \citenamefont
  {Lookman}}]{xue2016accelerated}%
  \BibitemOpen
  \bibfield  {author} {\bibinfo {author} {\bibfnamefont {D.}~\bibnamefont
  {Xue}}, \bibinfo {author} {\bibfnamefont {P.~V.}\ \bibnamefont
  {Balachandran}}, \bibinfo {author} {\bibfnamefont {J.}~\bibnamefont
  {Hogden}}, \bibinfo {author} {\bibfnamefont {J.}~\bibnamefont {Theiler}},
  \bibinfo {author} {\bibfnamefont {D.}~\bibnamefont {Xue}}, \ and\ \bibinfo
  {author} {\bibfnamefont {T.}~\bibnamefont {Lookman}},\ }\href@noop {}
  {\bibfield  {journal} {\bibinfo  {journal} {Nat. Commun.}\ }\textbf {\bibinfo
  {volume} {7}} (\bibinfo {year} {2016})}\BibitemShut {NoStop}%
\bibitem [{\citenamefont {Kiyohara}\ \emph {et~al.}(2016)\citenamefont
  {Kiyohara}, \citenamefont {Oda}, \citenamefont {Miyata},\ and\ \citenamefont
  {Mizoguchi}}]{Kiyoharae1600746}%
  \BibitemOpen
  \bibfield  {author} {\bibinfo {author} {\bibfnamefont {S.}~\bibnamefont
  {Kiyohara}}, \bibinfo {author} {\bibfnamefont {H.}~\bibnamefont {Oda}},
  \bibinfo {author} {\bibfnamefont {T.}~\bibnamefont {Miyata}}, \ and\ \bibinfo
  {author} {\bibfnamefont {T.}~\bibnamefont {Mizoguchi}},\ }\href {\doibase
  10.1126/sciadv.1600746} {\bibfield  {journal} {\bibinfo  {journal} {Science
  Adv.}\ }\textbf {\bibinfo {volume} {2}} (\bibinfo {year} {2016}),\
  10.1126/sciadv.1600746}\BibitemShut {NoStop}%
\bibitem [{\citenamefont {Seko}\ \emph {et~al.}(2017)\citenamefont {Seko},
  \citenamefont {Hayashi}, \citenamefont {Nakayama}, \citenamefont
  {Takahashi},\ and\ \citenamefont {Tanaka}}]{PhysRevB.95.144110}%
  \BibitemOpen
  \bibfield  {author} {\bibinfo {author} {\bibfnamefont {A.}~\bibnamefont
  {Seko}}, \bibinfo {author} {\bibfnamefont {H.}~\bibnamefont {Hayashi}},
  \bibinfo {author} {\bibfnamefont {K.}~\bibnamefont {Nakayama}}, \bibinfo
  {author} {\bibfnamefont {A.}~\bibnamefont {Takahashi}}, \ and\ \bibinfo
  {author} {\bibfnamefont {I.}~\bibnamefont {Tanaka}},\ }\href {\doibase
  10.1103/PhysRevB.95.144110} {\bibfield  {journal} {\bibinfo  {journal} {Phys.
  Rev. B}\ }\textbf {\bibinfo {volume} {95}},\ \bibinfo {pages} {144110}
  (\bibinfo {year} {2017})}\BibitemShut {NoStop}%
\bibitem [{\citenamefont {{Todorovi{\'c}}}\ \emph {et~al.}(2017)\citenamefont
  {{Todorovi{\'c}}}, \citenamefont {{Gutmann}}, \citenamefont {{Corander}},\
  and\ \citenamefont {{Rinke}}}]{2017arXiv170809274T}%
  \BibitemOpen
  \bibfield  {author} {\bibinfo {author} {\bibfnamefont {M.}~\bibnamefont
  {{Todorovi{\'c}}}}, \bibinfo {author} {\bibfnamefont {M.~U.}\ \bibnamefont
  {{Gutmann}}}, \bibinfo {author} {\bibfnamefont {J.}~\bibnamefont
  {{Corander}}}, \ and\ \bibinfo {author} {\bibfnamefont {P.}~\bibnamefont
  {{Rinke}}},\ }\href@noop {} {\bibfield  {journal} {\bibinfo  {journal} {ArXiv
  e-prints}\ } (\bibinfo {year} {2017})},\ \Eprint
  {http://arxiv.org/abs/1708.09274} {arXiv:1708.09274 [cond-mat.mtrl-sci]}
  \BibitemShut {NoStop}%
\bibitem [{\citenamefont {Pozun}\ \emph {et~al.}(2012)\citenamefont {Pozun},
  \citenamefont {Hansen}, \citenamefont {Sheppard}, \citenamefont {Rupp},
  \citenamefont {M{\"u}ller},\ and\ \citenamefont
  {Henkelman}}]{pozun2012optimizing}%
  \BibitemOpen
  \bibfield  {author} {\bibinfo {author} {\bibfnamefont {Z.~D.}\ \bibnamefont
  {Pozun}}, \bibinfo {author} {\bibfnamefont {K.}~\bibnamefont {Hansen}},
  \bibinfo {author} {\bibfnamefont {D.}~\bibnamefont {Sheppard}}, \bibinfo
  {author} {\bibfnamefont {M.}~\bibnamefont {Rupp}}, \bibinfo {author}
  {\bibfnamefont {K.-R.}\ \bibnamefont {M{\"u}ller}}, \ and\ \bibinfo {author}
  {\bibfnamefont {G.}~\bibnamefont {Henkelman}},\ }\href@noop {} {\bibfield
  {journal} {\bibinfo  {journal} {The Journal of chemical physics}\ }\textbf
  {\bibinfo {volume} {136}},\ \bibinfo {pages} {174101} (\bibinfo {year}
  {2012})}\BibitemShut {NoStop}%
\bibitem [{\citenamefont {Iwahara}\ \emph {et~al.}(1993)\citenamefont
  {Iwahara}, \citenamefont {Yajima}, \citenamefont {Hibino}, \citenamefont
  {Ozaki},\ and\ \citenamefont {Suzuki}}]{iwahara1993protonic}%
  \BibitemOpen
  \bibfield  {author} {\bibinfo {author} {\bibfnamefont {H.}~\bibnamefont
  {Iwahara}}, \bibinfo {author} {\bibfnamefont {T.}~\bibnamefont {Yajima}},
  \bibinfo {author} {\bibfnamefont {T.}~\bibnamefont {Hibino}}, \bibinfo
  {author} {\bibfnamefont {K.}~\bibnamefont {Ozaki}}, \ and\ \bibinfo {author}
  {\bibfnamefont {H.}~\bibnamefont {Suzuki}},\ }\href@noop {} {\bibfield
  {journal} {\bibinfo  {journal} {Solid State Ionics}\ }\textbf {\bibinfo
  {volume} {61}},\ \bibinfo {pages} {65} (\bibinfo {year} {1993})}\BibitemShut
  {NoStop}%
\bibitem [{\citenamefont {M{\"u}nch}\ \emph {et~al.}(2000)\citenamefont
  {M{\"u}nch}, \citenamefont {Kreuer}, \citenamefont {Seifert},\ and\
  \citenamefont {Maier}}]{munch2000proton}%
  \BibitemOpen
  \bibfield  {author} {\bibinfo {author} {\bibfnamefont {W.}~\bibnamefont
  {M{\"u}nch}}, \bibinfo {author} {\bibfnamefont {K.-D.}\ \bibnamefont
  {Kreuer}}, \bibinfo {author} {\bibfnamefont {G.}~\bibnamefont {Seifert}}, \
  and\ \bibinfo {author} {\bibfnamefont {J.}~\bibnamefont {Maier}},\
  }\href@noop {} {\bibfield  {journal} {\bibinfo  {journal} {Solid State
  Ionics}\ }\textbf {\bibinfo {volume} {136}},\ \bibinfo {pages} {183}
  (\bibinfo {year} {2000})}\BibitemShut {NoStop}%
\bibitem [{\citenamefont {Bj{\"o}rketun}\ \emph {et~al.}(2007)\citenamefont
  {Bj{\"o}rketun}, \citenamefont {Sundell},\ and\ \citenamefont
  {Wahnstr{\"o}m}}]{bjorketun2007effect}%
  \BibitemOpen
  \bibfield  {author} {\bibinfo {author} {\bibfnamefont {M.~E.}\ \bibnamefont
  {Bj{\"o}rketun}}, \bibinfo {author} {\bibfnamefont {P.~G.}\ \bibnamefont
  {Sundell}}, \ and\ \bibinfo {author} {\bibfnamefont {G.}~\bibnamefont
  {Wahnstr{\"o}m}},\ }\href@noop {} {\bibfield  {journal} {\bibinfo  {journal}
  {Physical Review B}\ }\textbf {\bibinfo {volume} {76}},\ \bibinfo {pages}
  {054307} (\bibinfo {year} {2007})}\BibitemShut {NoStop}%
\bibitem [{\citenamefont {Sundell}\ \emph {et~al.}(2007)\citenamefont
  {Sundell}, \citenamefont {Bj{\"o}rketun},\ and\ \citenamefont
  {Wahnstr{\"o}m}}]{sundell2007density}%
  \BibitemOpen
  \bibfield  {author} {\bibinfo {author} {\bibfnamefont {P.~G.}\ \bibnamefont
  {Sundell}}, \bibinfo {author} {\bibfnamefont {M.~E.}\ \bibnamefont
  {Bj{\"o}rketun}}, \ and\ \bibinfo {author} {\bibfnamefont {G.}~\bibnamefont
  {Wahnstr{\"o}m}},\ }\href@noop {} {\bibfield  {journal} {\bibinfo  {journal}
  {Physical Review B}\ }\textbf {\bibinfo {volume} {76}},\ \bibinfo {pages}
  {094301} (\bibinfo {year} {2007})}\BibitemShut {NoStop}%
\bibitem [{\citenamefont {Haugsrud}\ and\ \citenamefont
  {Norby}(2006{\natexlab{a}})}]{haugsrud2006proton1}%
  \BibitemOpen
  \bibfield  {author} {\bibinfo {author} {\bibfnamefont {R.}~\bibnamefont
  {Haugsrud}}\ and\ \bibinfo {author} {\bibfnamefont {T.}~\bibnamefont
  {Norby}},\ }\href@noop {} {\bibfield  {journal} {\bibinfo  {journal} {Nature
  Materials}\ }\textbf {\bibinfo {volume} {5}},\ \bibinfo {pages} {193}
  (\bibinfo {year} {2006}{\natexlab{a}})}\BibitemShut {NoStop}%
\bibitem [{\citenamefont {Haugsrud}\ and\ \citenamefont
  {Norby}(2006{\natexlab{b}})}]{haugsrud2006proton2}%
  \BibitemOpen
  \bibfield  {author} {\bibinfo {author} {\bibfnamefont {R.}~\bibnamefont
  {Haugsrud}}\ and\ \bibinfo {author} {\bibfnamefont {T.}~\bibnamefont
  {Norby}},\ }\href@noop {} {\bibfield  {journal} {\bibinfo  {journal} {Nature
  Materials}\ }\textbf {\bibinfo {volume} {5}},\ \bibinfo {pages} {193}
  (\bibinfo {year} {2006}{\natexlab{b}})}\BibitemShut {NoStop}%
\bibitem [{\citenamefont {Bl{\"o}chl}(1994)}]{blochl1994projector}%
  \BibitemOpen
  \bibfield  {author} {\bibinfo {author} {\bibfnamefont {P.~E.}\ \bibnamefont
  {Bl{\"o}chl}},\ }\href@noop {} {\bibfield  {journal} {\bibinfo  {journal}
  {Physical Review B}\ }\textbf {\bibinfo {volume} {50}},\ \bibinfo {pages}
  {17953} (\bibinfo {year} {1994})}\BibitemShut {NoStop}%
\bibitem [{\citenamefont {Kresse}\ and\ \citenamefont
  {Hafner}(1993)}]{kresse1993ab}%
  \BibitemOpen
  \bibfield  {author} {\bibinfo {author} {\bibfnamefont {G.}~\bibnamefont
  {Kresse}}\ and\ \bibinfo {author} {\bibfnamefont {J.}~\bibnamefont
  {Hafner}},\ }\href@noop {} {\bibfield  {journal} {\bibinfo  {journal}
  {Physical Review B}\ }\textbf {\bibinfo {volume} {48}},\ \bibinfo {pages}
  {13115} (\bibinfo {year} {1993})}\BibitemShut {NoStop}%
\bibitem [{\citenamefont {Kresse}\ and\ \citenamefont
  {Furthm{\"u}ller}(1996)}]{kresse1996efficiency}%
  \BibitemOpen
  \bibfield  {author} {\bibinfo {author} {\bibfnamefont {G.}~\bibnamefont
  {Kresse}}\ and\ \bibinfo {author} {\bibfnamefont {J.}~\bibnamefont
  {Furthm{\"u}ller}},\ }\href@noop {} {\bibfield  {journal} {\bibinfo
  {journal} {Computational Materials Science}\ }\textbf {\bibinfo {volume}
  {6}},\ \bibinfo {pages} {15} (\bibinfo {year} {1996})}\BibitemShut {NoStop}%
\bibitem [{\citenamefont {Kresse}\ and\ \citenamefont
  {Joubert}(1999)}]{kresse1999ultrasoft}%
  \BibitemOpen
  \bibfield  {author} {\bibinfo {author} {\bibfnamefont {G.}~\bibnamefont
  {Kresse}}\ and\ \bibinfo {author} {\bibfnamefont {D.}~\bibnamefont
  {Joubert}},\ }\href@noop {} {\bibfield  {journal} {\bibinfo  {journal}
  {Physical Review B}\ }\textbf {\bibinfo {volume} {59}},\ \bibinfo {pages}
  {1758} (\bibinfo {year} {1999})}\BibitemShut {NoStop}%
\bibitem [{\citenamefont {Fjeld}\ \emph {et~al.}(2010)\citenamefont {Fjeld},
  \citenamefont {Toyoura}, \citenamefont {Haugsrud},\ and\ \citenamefont
  {Norby}}]{fjeld2010proton}%
  \BibitemOpen
  \bibfield  {author} {\bibinfo {author} {\bibfnamefont {H.}~\bibnamefont
  {Fjeld}}, \bibinfo {author} {\bibfnamefont {K.}~\bibnamefont {Toyoura}},
  \bibinfo {author} {\bibfnamefont {R.}~\bibnamefont {Haugsrud}}, \ and\
  \bibinfo {author} {\bibfnamefont {T.}~\bibnamefont {Norby}},\ }\href@noop {}
  {\bibfield  {journal} {\bibinfo  {journal} {Physical Chemistry Chemical
  Physics}\ }\textbf {\bibinfo {volume} {12}},\ \bibinfo {pages} {10313}
  (\bibinfo {year} {2010})}\BibitemShut {NoStop}%
\bibitem [{Note1()}]{Note1}%
  \BibitemOpen
  \bibinfo {note} {The 1D long-range paths along the $a$- and $b$-axes are
  alternatively arranged in the $c$-axis direction due to the four-fold screw
  symmetry. Note that the broken lines in Fig. \ref {fig:result2}(b) show only
  the paths along the $b$-axis.}\BibitemShut {Stop}%
\end{thebibliography}
\end{document}